\documentclass [useAMS]{mn2e}
\usepackage {graphicx}
\begin{document}

\title{Effects of Magnetic Coupling on Radiation from Accretion Disc around a Kerr Black Hole}
\author[]{Zhao-Ming Gan$^{1}$, Ding-Xiong Wang$^{1,2}$ and Yang Li$^{1}$\\\\
$1$ Department of Physics, Huazhong University of Science and Technology, Wuhan 430074, China\\
$2$ Send offprint requests to: D.-X. Wang(dxwang@hust.edu.cn)}

\maketitle

\begin{abstract}
The effects of magnetic coupling (MC) process on the inner edge of
the disc are discussed in detail. It is shown that the inner edge
can deviate from the innermost stable circular orbit (ISCO) due to
the magnetic transfer of energy and angular momentum between a
Kerr black hole (BH) and its surrounding accretion disc. It turns
out that the inner edge could move inward and outward for the BH
spin $a_{*}$ being greater and less than 0.3594, respectively. The
MC effects on disc radiation are discussed based on the displaced
inner edge. A very steep emissivity can be provided by the MC
process, which is consistent with the observation of MCG-6-30-15.
In addition, the BH spins of GRO J1655-40 and GRS 1915+105 are
detected by X-ray continuum fitting based on this model.
\end{abstract}


\section{INTRODUCTION}
As is well shown, disc accretion is a very efficient energy source
in astrophysics, which is widely used to explain the high-energy
radiation of AGNs, X-ray binaries and so on. In the model of
standard accretion disc (SAD), the motion of the accreting matter
is assumed to be Keplerian with a small radial velocity, and the
inner edge of the disc lies at the innermost stable circular orbit
(ISCO) of radius $r_{ISCO} $, within which the matter plunges very
fast onto the black hole (BH). In this case, there would be no
significant torques exerted at the inner edge, and the ``no torque
boundary condition'' is a very good approximation (Bardeen 1970,
Shakura \& Sunyaev 1973, Novikov \& Thorne 1973, Page \& Thorne
1974).

Not long ago, some authors discussed the possibility that magnetic
fields may induce a nonzero torque at ISCO, which involves the
magnetic connection of the matter in the plunging region inside
ISCO with the matter in the disc outside ISCO (Krolik 1999; Gammie
1999; Agol \& Krolik 2000). On the other hand, some authors argued
that the torque at ISCO of a thin disk is very weak, since it
depends on the vertical thickness of the disk (Paczynski 2000; Li
2002a, hereafter L02; Afshordi \& Paczynski 2003).

Recently, much attention has been paid to the magnetic coupling
(MC) between a BH and its surrounding accretion disc (Blandford
1999, Li 2002a, hereafter L02, Wang et al. 2002), which can be
regarded as one of the variants of the Blandford-Znajek (BZ)
process (Blandford \& Znajek 1977). By virtue of the magnetic
field connecting the rotating BH and the disc, energy and angular
momentum are transferred between the BH and the disc, just like
the energy transportation between a dynamo and a motor (Macdonald
\& Thorne 1982, Thorne, Price \& MacDonald 1986). In other words,
the BH exerts a torque at the disc in the MC process, resulting in
the transfer of energy and angular momentum. The transfer
direction is determined by the difference between the rotation of
the BH and that of the disc, i.e., if the BH rotates faster than
the disc, energy and angular momentum are extracted from the BH
and transferred to the disc, otherwise the direction is reversed.

Recently, Li (2004, hereafter L04) studied the MC between a BH and
its surrounding disc. It is argued that the inner edge of the disc
moves out to a radius where the angular velocity of the disc is
equal to that of the BH for $a_{*}<0.3594$. However, the inner
edge remains at ISCO for the BH spin with $a_{*}>0.3594$. This
result seems somewhat inconsistent. Why is the inner edge of the
disc affected by the MC for $a_{*}<0.3594$, but not for
$a_{*}>0.3594$ ?

In this paper, we intend to discuss the MC effects on the inner
edge of the disc. In our model the inner edge of the disc is not
assumed to be ISCO in advance. The MC effects on the disc
radiation lie in the following aspects. The profile of the
radiation flux and the interior viscous torque at the disc can be
changed due to the magnetic torque exerted at the disc. It is
shown that if the BH rotates faster than the disc, the disc
radiation would be more concentrated at the inner disc in the MC
process. This feature can be used to explain the very steep
emissivity index in the inner disc, which is consistent with the
\textit{XMM-Newton} observation of the nearby bright Seyfert 1
galaxy MCG-6-30-15 (Wilms et al. 2001; Li 2002b; Wang et al. 2003
hereafter W03).

The disc radiation is derived by considering the magnetic transfer
of the energy and angular momentum between a rotating BH and a
relativistic thin disc (Page \& Thorne 1974; L02), and a criterion
of determining the inner edge of the disc is proposed based on a
reasonable constraint, i.e., a reasonable disc radiation should
not be negative. It is shown that the value $a_{*}=0.3594$ can be
regarded as a critical BH spin, which corresponds to the angular
velocity of ISCO equal to that of the BH. As the BH spin $a_{*}$
is less than 0.3594, i.e., the energy and angular momentum are
transferred from the inner disc to the BH, and the balance of
energy and angular momentum on the disc would be disrupted in the
MC process. Thus the inner disc becomes unstable, as the disc
cannot suffer a negative interior viscous torque. It turns out
that the MC effects on the disc radiation are so strong that the
inner edge of the disc has to move outward and inward from ISCO
for the BH spin $a_{*}$ less and greater than 0.3594,
respectively.

In order to facilitate the discussion of the MC effects on the
inner edge of the disc we make the following assumptions:

1.The large scale magnetic field remains constant at the BH
horizon, while it varies as a power-law with the disc radius as
given in W03 rather than concentrated at ISCO.

2. The disc is thin, perfectly conducting and Keplerian, lying in
the equatorial plane of a Kerr BH;

3. Considering that the ``no torque boundary condition'' remains
controversial, we assume that no torque is exerted at the inner
edge of a thin disc in the MC with a central BH.

This paper is organized as follows. In \S 2 the properties of the
inner edge of SAD and the radial profiles of specific energy and
angular momentum are discussed in detail. In \S 3 a criterion for
the inner edge of a disc is proposed by considering the MC
effects. In \S 4 some of the characteristics and applications of
the MC process are presented, including fitting the steep
emissivity index of MCG-6-30-15 and the BH spins of GRO J1655-40
and GRS 1915+105. Finally, in \S 5, we summarize our main results.
Throughout this paper Boyer-Lindquist coordinates
$(t,r,\theta,\varphi)$ and the geometric units $G = c = 1$ are
used.

\section{INNER EDGE OF SAD WITH NO TORQUE BOUNDARY CONDITION}

As is well known, the full description of a Kerr BH needs only two
parameters, i.e., mass $M$ and spin $a_{*}\equiv a/M \equiv
J/M^{2}$ ($-1<a_{*}< 1$).

Based on Page \& Thorne (1974) and L02 we derive the expressions
for the radiation flux, interior viscous torque and total
luminosity of a relativistic thin disc as follows.

\begin{equation}
\label{eq1}
\begin{array}{l}
F_{DA} = {\frac{{1}}{{4\pi r}}}{\frac{{ - d\Omega _{D} /
dr}}{{(E^{\dag}  - \Omega _{D} L^{\dag} )^{2}}}}[{\int_{r_{in}}
^{r} {(E^{\dag}  - \Omega _{D} L^{\dag} )}} \dot {M}_{D}
{\frac{{dL^{\dag} }}{{dr}}}dr \\\\
\quad \quad + g_{in} (E_{in}^{\dag}  - \Omega _{in}
L_{in}^{\dag})]
\end{array}
\end{equation}

\begin{equation}
\label{eq2}
\begin{array}{l}

g_{DA} = {\frac{{E^{\dag}  - \Omega _{D} L^{\dag} }}{{ - d\Omega
_{D} / dr}}}4\pi r \cdot F_{DA}
\end{array}
\end{equation}

\begin{equation}
\label{eq3} {\cal L}_{DA} = 2{\int_{r_{in}}^{\infty}{E^{\dag
}F_{DA}\cdot} }2\pi r \cdot dr =\dot{M}_{D}(1-E_{in}^{^{\dag}} )
+g_{in}\cdot\Omega_{in}
\end{equation}

\noindent where $\dot {M}_{D}$ is the accretion rate. The
quantities $F_{DA} $ and $g_{DA} $ are the radiation flux and
interior viscous torque of the disc, respectively. It is
emphasized that these quantities can never be negative in any
physical cases.  $E^{\dag}$, $L^{\dag}$ and $\Omega_{D}$ are
respectively the specific energy, specific angular momentum and
angular velocity of the test particles moving along geodesic
circular orbits in the equatorial plane of a Kerr BH (Bardeen,
Press \& Teukolsky 1972, hereafter B72).

The quantity ${\cal L}_{DA} $ is the total luminosity, and $g_{in}
$ is the exterior torque exerted at the inner boundary with
angular velocity $\Omega_{in}=\Omega_{D}(r_{in})$. Generally, the
``no torque boundary condition'' is assumed in the theory of SAD,
i.e., $g_{in}=0$ (Page \& Thorne 1974). From equation (1)-(3) we
find that the radiation flux and interior viscous torque could be
significantly different in the case of $g_{in} \ne 0$, by which an
extra energy is provided.

\begin{figure}
\vspace{0.5cm}
\begin{center}
{\includegraphics[width=5.8cm]{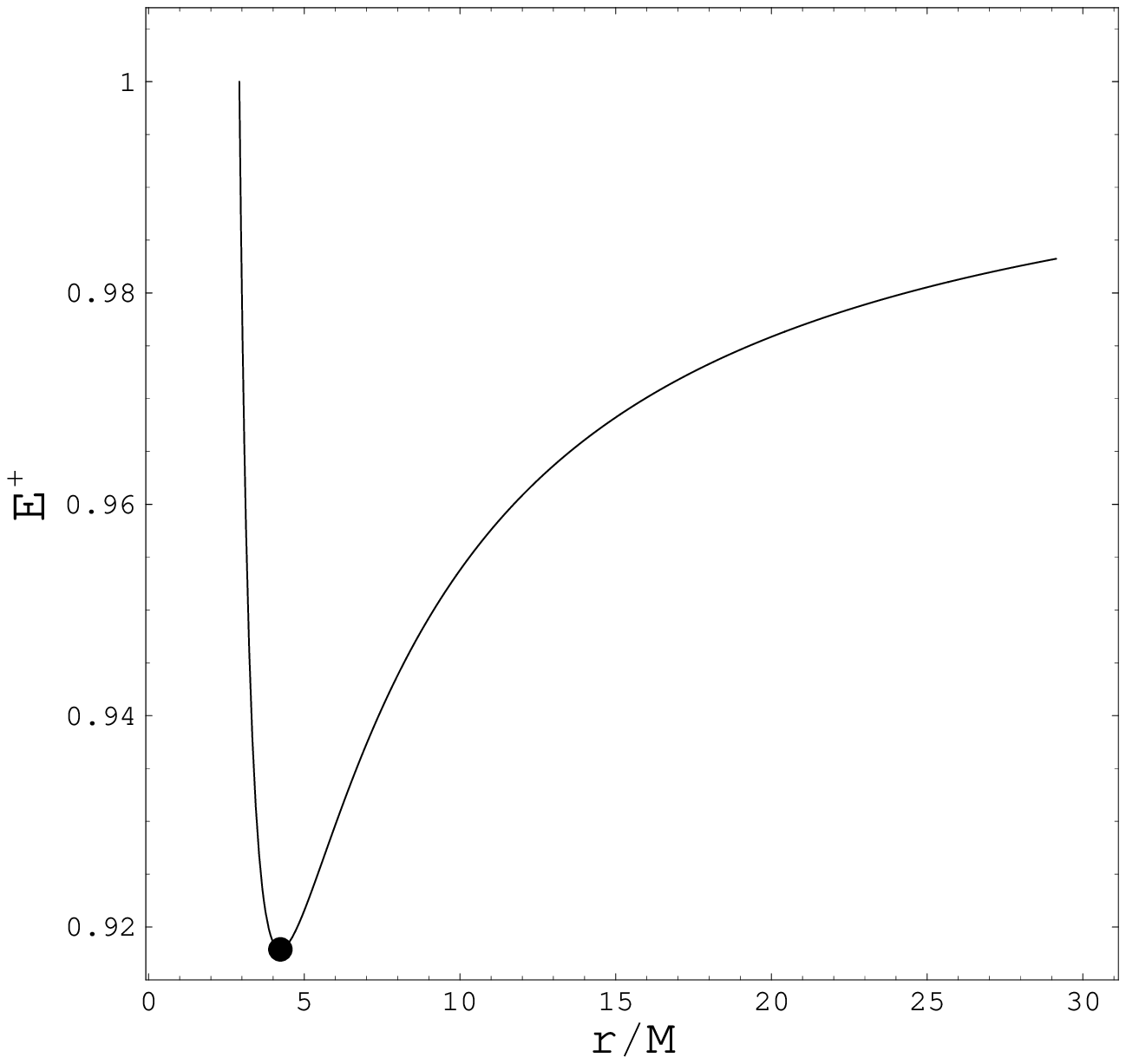} \hfill
\includegraphics[width=5.7cm]{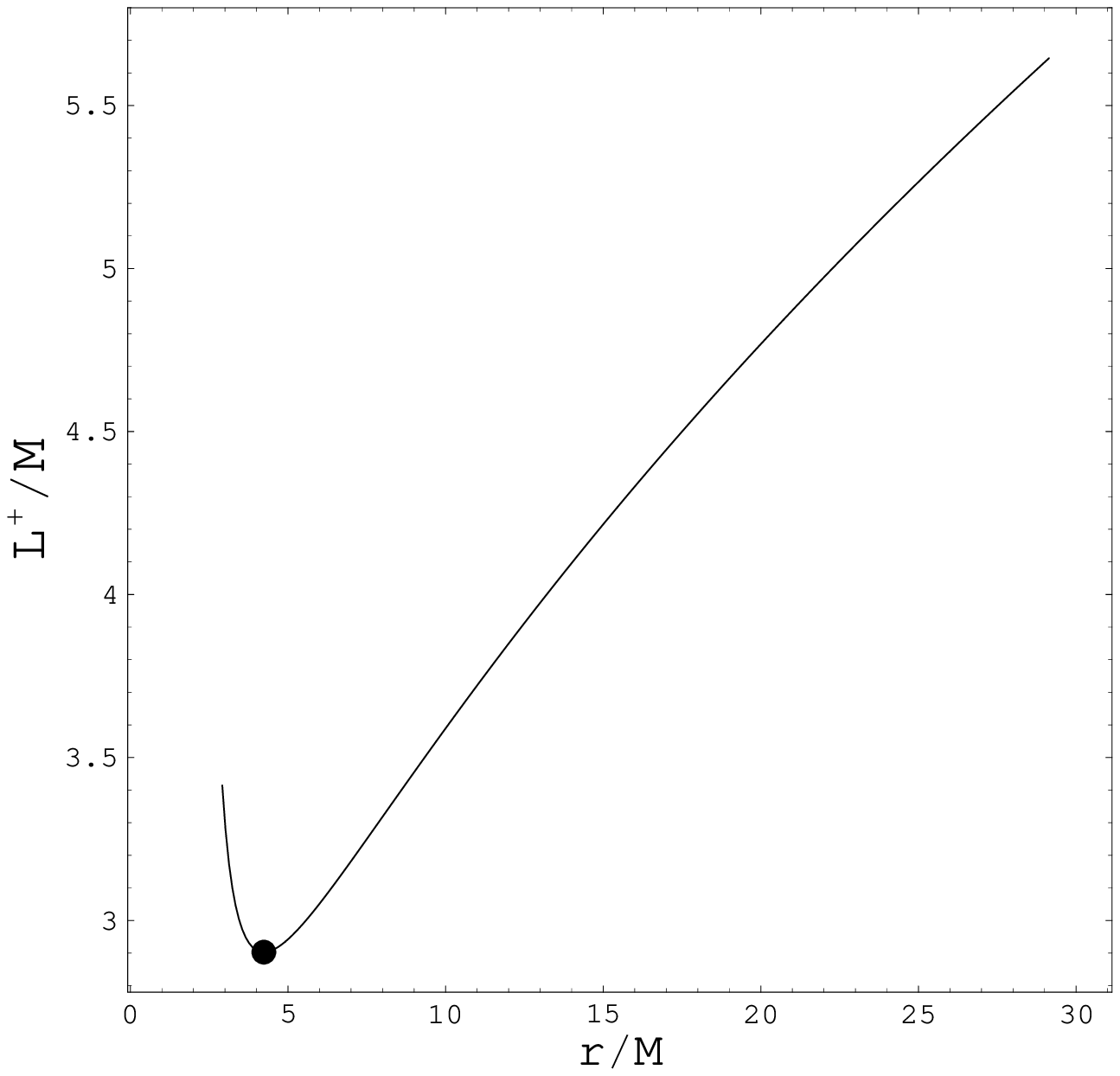}}
\caption{Specific energy and angular momentum of the accreting
matter of SAD versus the disc radius with $a_{*}$=0.5. The black
points indicate the position radius of ISCO.} \label{fig1}
\end{center}
\end{figure}

\begin{figure*}
\vspace{0.5cm}
\begin{center}
{\includegraphics[width=5.4cm]{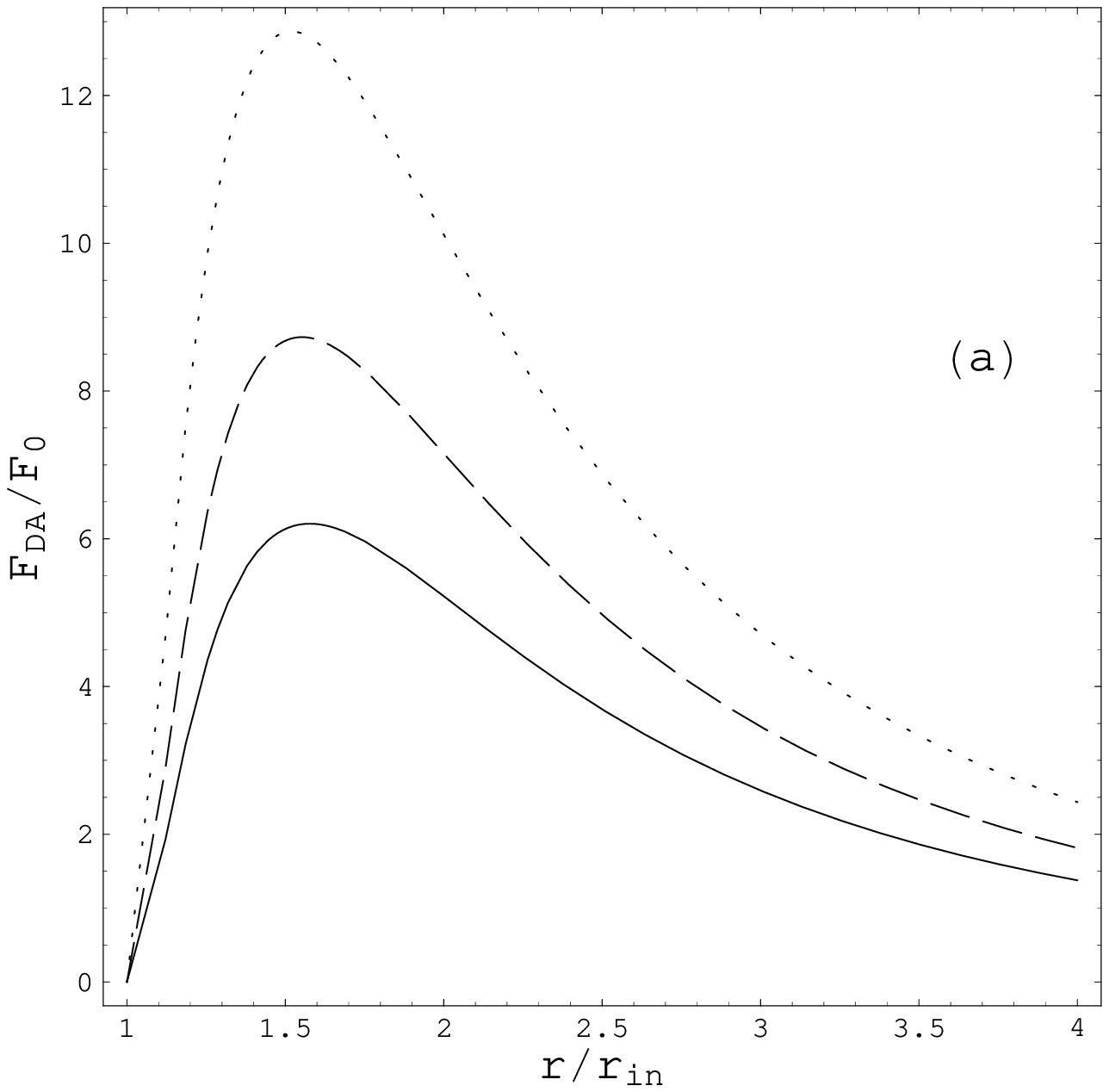}\hfill
\includegraphics[width=5.4cm]{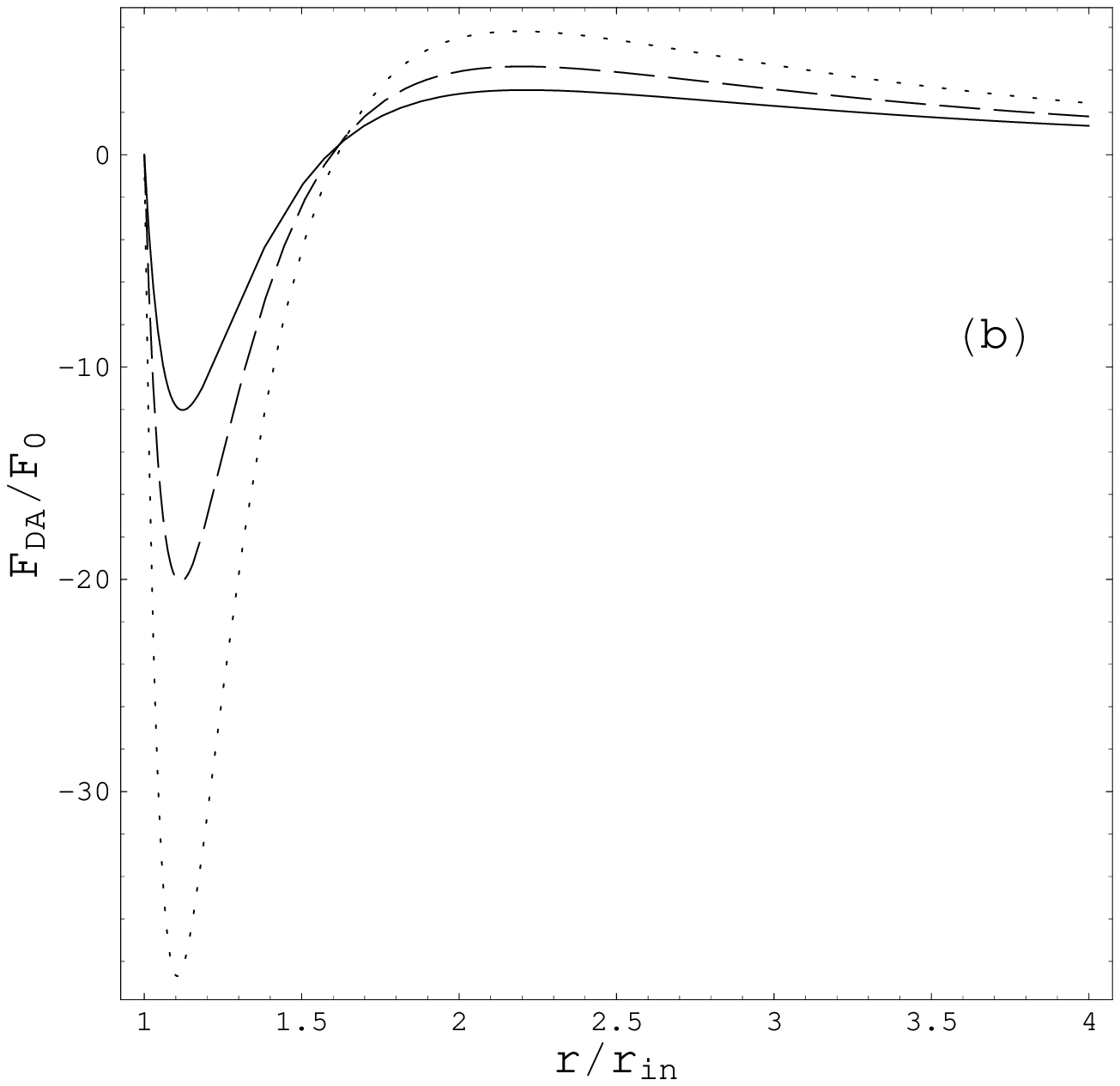}\hfill
\includegraphics[width=5.4cm]{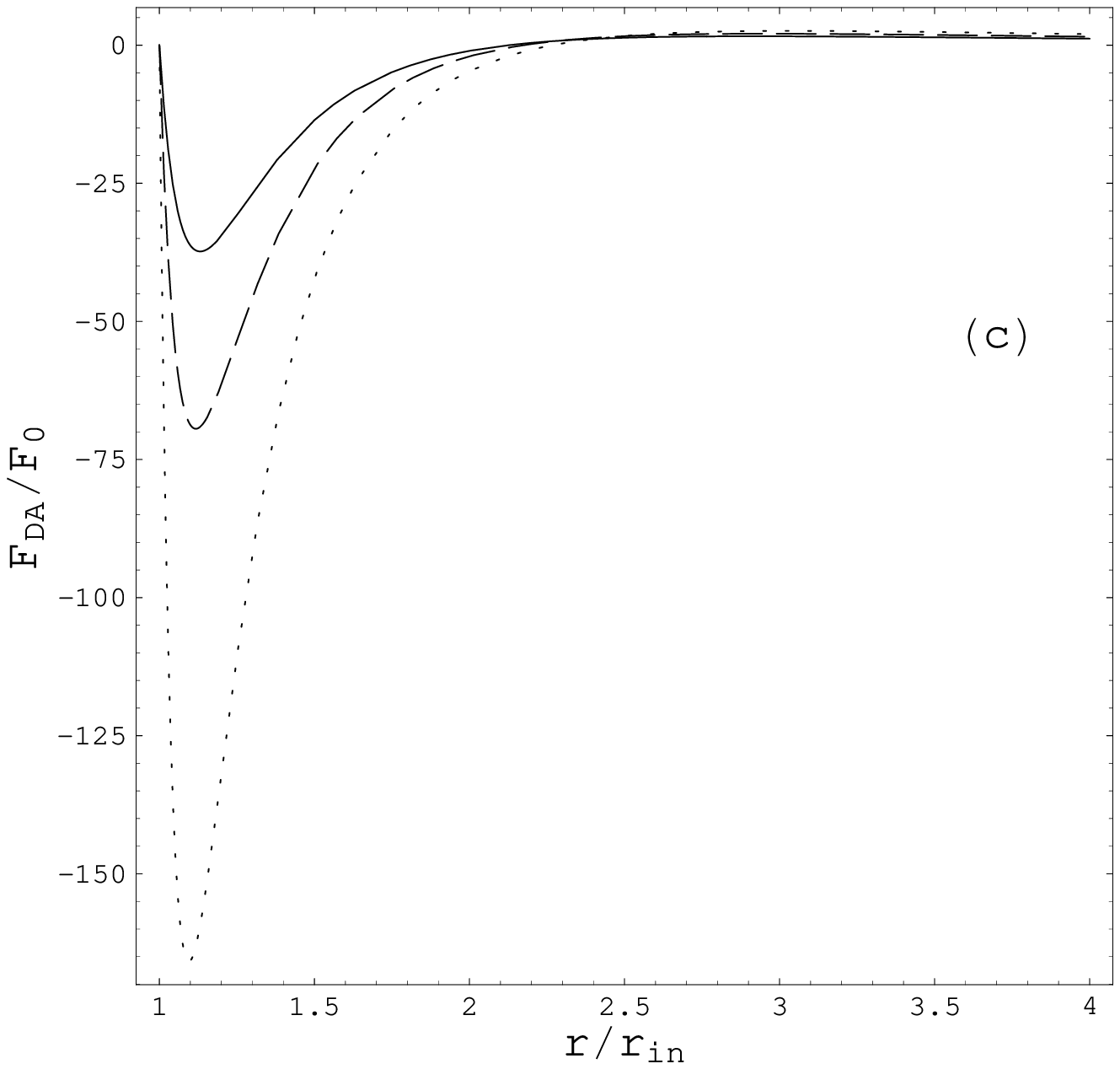}\hfill}
\caption{Radiation flux from SAD with different inner edges versus
the disc radius for different BH spins, where $a_{*}=$0.8, 0.6 and
0.3 correspond to dotted, dashed and solid lines, respectively.
The parameter $\lambda=$1.0, 0.90 and 0.85 correspond to panels
(a), (b) and (c), respectively.} \label{fig2 }
\end{center}
\end{figure*}

The specific energy and angular momentum vary non-monotonically
with the disc radius, attaining their minima at ISCO indicated by
a black dot as shown in Figure 1. The radial profile of $E^{\dag}$
and $L^{\dag}$ is very different from the monotonous profile of
these quantities obtained in Newtonian mechanics. It is the radial
feature of $E^{\dag}$ and $L^{\dag}$ in general relativity that
results in two constraints to the inner edge of SAD: (1) it is
always located at or outside ISCO without extra energy and angular
momentum transferred to the accreting matter, and (2) it could
move inward within ISCO, provided that some extra energy and
angular momentum are transferred to the accreting matter. In \S 3
we shall elaborate this point in detail, and propose a criterion
for the inner edge of a thin disc by considering the MC effects.

The lower limit $r_{in}$ to the integral in equations (1) and (3)
is the radius of the inner edge of the disc, which is usually
assumed to lie at ISCO in SAD. Abramowitz et al. (1978) argued
that if the disc is geometrically thick, its inner edge locates
between ISCO and the innermost bound circular orbit.

 A parameter $\lambda $ is introduced to label the position of the inner edge,
which is defined as

\begin{equation}
\label{eq4} r_{in} \equiv \lambda^{2} \ r_{ISCO},\quad \quad for
\quad \lambda \ge  \sqrt{r_{H}/r_{ISCO}},
\end{equation}

\noindent where $r_{H}$ is the radius of the BH horizon (B72).

Combining equations (1) with (4), we have the curves of
$F_{DA}/F_{0}$ versus $r/r_{in}$ for $g_{in}=0$ as shown in Figure
2, where the quantity $F_{0} $ is defined as $F_{0}= 10^{-4} \cdot
\dot{M}_{D}/r_{in}^{2}$.

As shown in Figure 2a, the radiation flux becomes zero at
$r_{ISCO}$ with $\lambda = 1$. However, an unphysical disc region
with negative radiation flux appears for $\lambda $ less than
unity as shown in Figures 2b and 2c, implying that the radius
$r_{in}$ cannot be smaller than $r_{ISCO}$. This unphysical case
can be understood based on the radial profile of $E^{\dag}$,
$L^{\dag}$ given in Figure 1.

In SAD the accreting matter outside $r_{ISCO}$ losses its energy
and angular momentum due to the differential rotation, and thus
enters into the smaller circular orbit. Since the radius
$r_{ISCO}$ corresponds to the minimum specific energy and specific
angular momentum as shown in Figure 1, the accreting matter has to
absorb rather than release energy and angular momentum to get into
the smaller orbit. The real picture inside $r_{ISCO}$ is that the
accreting matter is plunged into the BH with a huge radial
velocity due to lack of enough angular momentum.

Thus we conclude that SAD with an inner edge within $r_{ISCO} $ is
not stable, while any radius larger than $r_{ISCO} $ is possible.
It is worth noting that $r_{ISCO} $, the radius of ISCO, is not
the unique choice for the inner edge of a standard thin disc, only
being the smallest radius among all of the possible ones. In
addition, as argued in the next section, the inner edge of an
accretion disc can deviate from ISCO in the MC process due to the
magnetic transfer of energy and angular momentum between a
rotating BH and its surrounding accretion disc.

\section{MC EFFECTS ON INNER EDGE OF DISC}

The magnetic field configuration of our model is shown in Figure
3. The parameter $n$ is the power-law index indicating the
variation of the poloidal magnetic field with the disc radius,
$B_{D}^{P} \propto r^{ - n}$. The mapping relation between between
the angular coordinate on the BH horizon and the radial coordinate
on the disc can be derived from the conservation of the magnetic
flux (Wang et al. 2002, W03) and it reads

\begin{figure}
\vspace{0.5cm}
\begin{center}
\includegraphics[width=8.0cm]{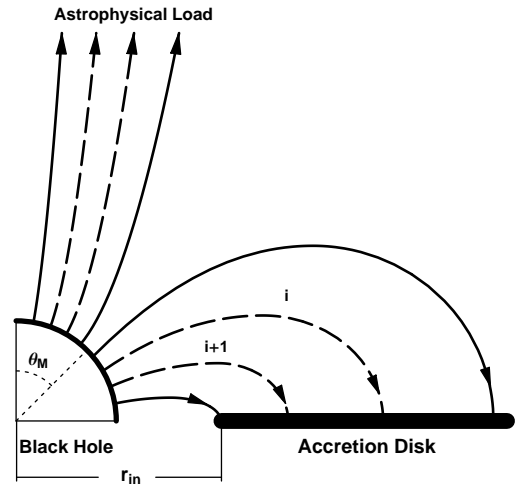}
\caption{The poloidal magnetic field configuration in our model.}
\label{fig3}
\end{center}
\end{figure}

\begin{figure*}
\vspace{0.5cm}
\begin{center}
{\includegraphics[width=5.4cm]{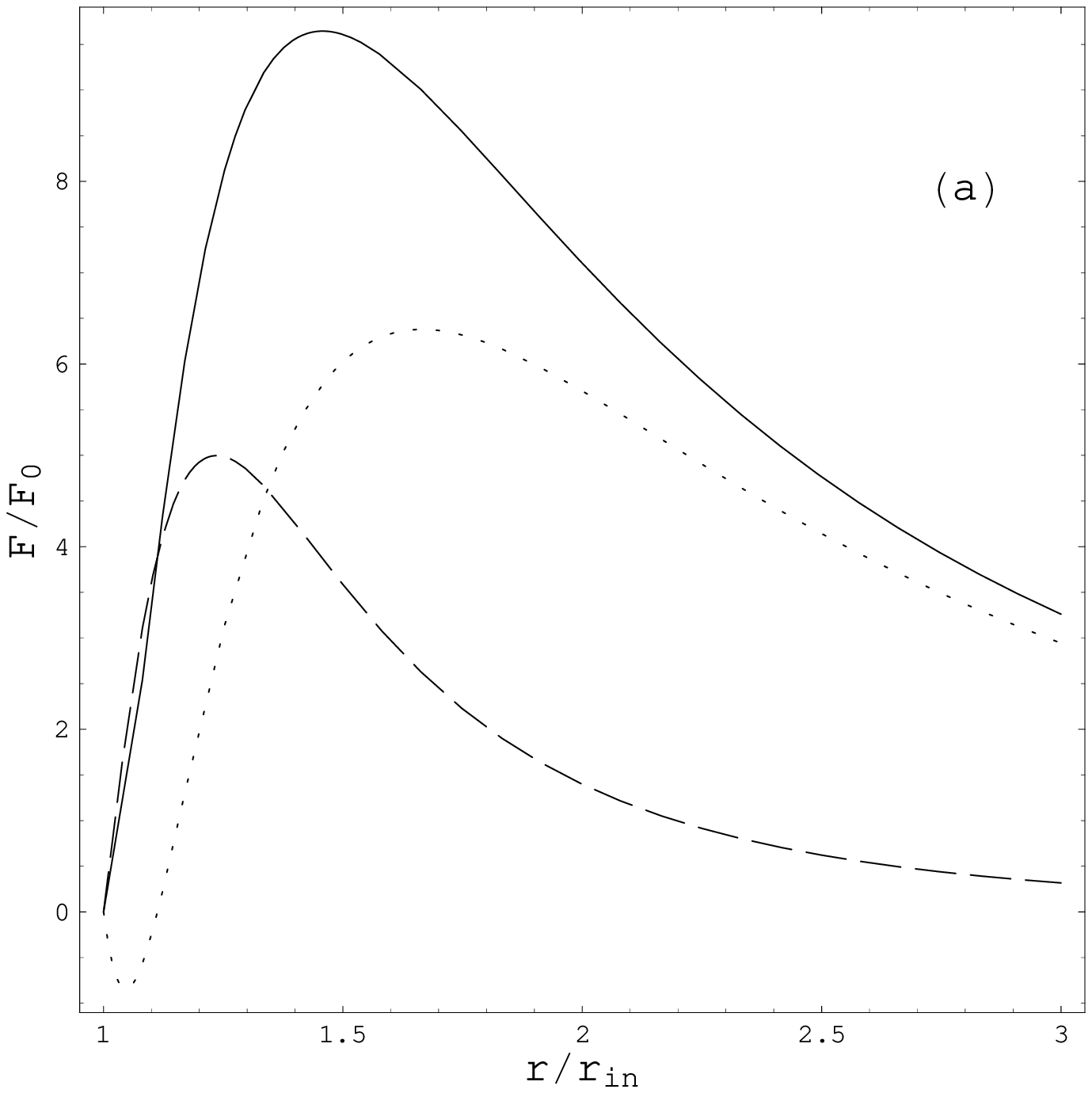}
\includegraphics[width=5.4cm]{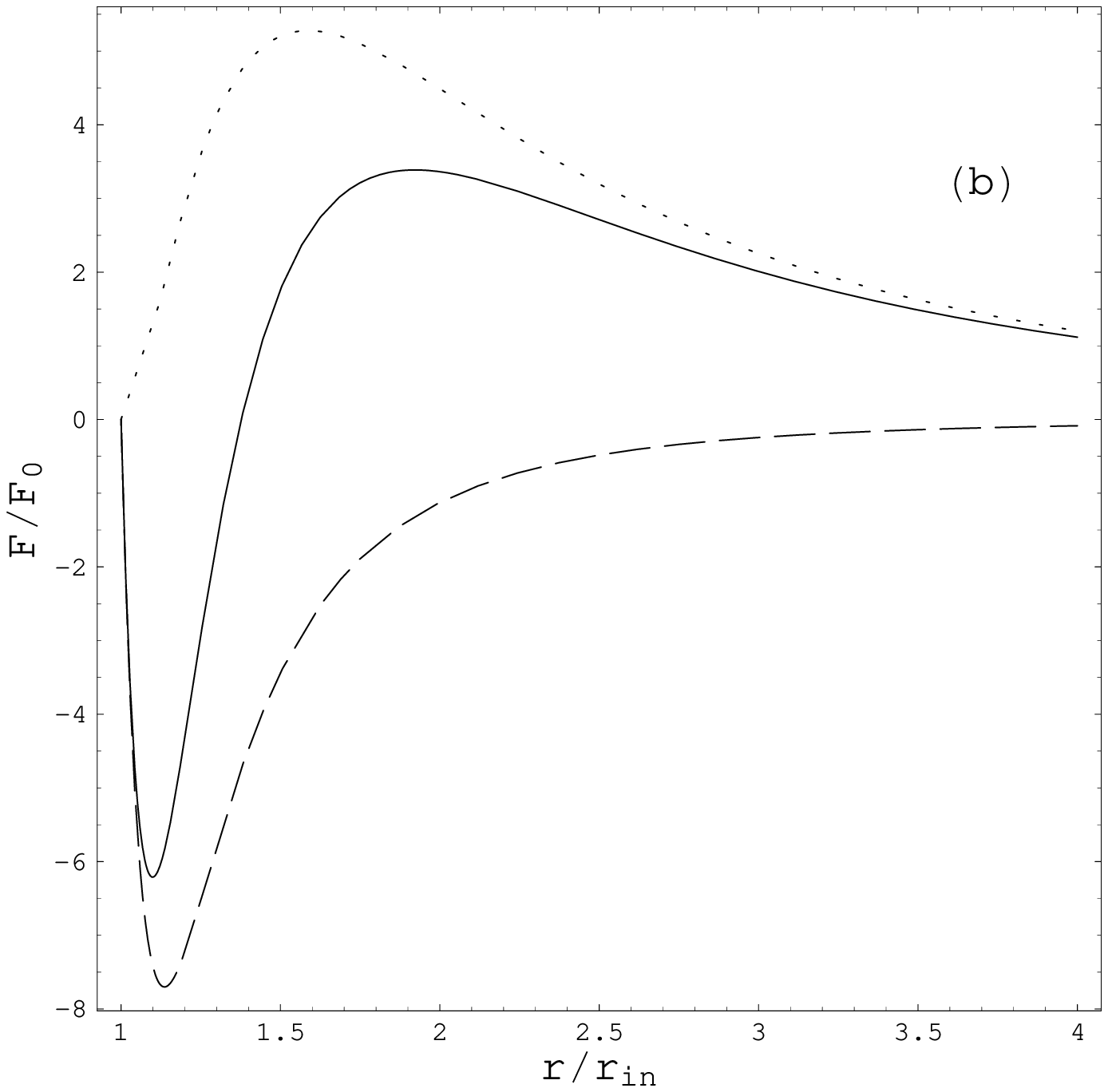}}
\caption{ The radiation fluxes versus $r/r_{in}$ with
$F_{total}/F_{0}$, $F_{DA}/F_{0}$ and $F_{MC}/F_{0}$ plotted in
solid, dotted and dashed lines, respectively. The parameters $n =
6$ and $\kappa _{m} = 1$ are taken with $\lambda=0.974$,
$a_{*}=0.45$ and $\lambda=1.000$, $a_{*}=0.10$, in panels (a) and
(b), respectively.} \label{fig4}
\end{center}
\end{figure*}

\begin{equation}
\label{eq5}
\begin{array}{l}
\cos \theta = {\int_{r_{in}}^{r}{{\rm G}({a_{ *}; {r}',n})\ d
{r}'}}  + C_{0},
\end{array}
\end{equation}

\begin{equation}
\label{eq6}
\begin{array}{l}
{\rm G}\left( {a_{ *}  ;r ,n} \right) =\\
\quad \quad {\frac{{\left(r/r_{in}\right)^{1-n} \sqrt {1 +
a_{*}^{2}M^{2} r^{-2} + 2a_{*}^{2}M^{3}r^{-3}}}}{{2\sqrt {\left(
{1 + a_{*}^{2} M^{2}r_{in}^{-2} +
2a_{*}^{2}M^{3}r_{in}^{-3}}\right)\left( {1 - 2Mr^{-1}+a_{*}
^{2}M^{2}r^{- 2}} \right)}}}}

\end{array}
\end{equation}

\noindent where $C_{0}$ is an integral constant (we adopt
$C_{0}=\cos 0.45\pi$ in this paper). According to the conservation
of the magnetic flux, $B_{D}^{P}$ and the poloidal magnetic field
on the horizon $B_{H}$ are related by (Wang et al. 2002, W03)

\begin{equation}
\label{eq7}
\begin{array}{l} d\psi=-B_{H}\cdot 2\pi(r_{H}^{2}+a^{2})\sin\theta \cdot d\theta\\
\ \ \ \ \ =B_{D}^{P}\cdot 2\pi
\sqrt{\frac{r^{4}+r^{2}a^{2}+2a^{2}Mr}{r^{2}+a^{2}-2Mr}} \cdot dr,
\end{array}
\end{equation}

\noindent where $\psi=\psi(r,\theta)$ is the magnetic flux through
a surface bounded by a circle with $r=constant$ and $r=constant$.

\

Based on Macdonald \& Thorne (1982) and W03 we express the MC
torque and power as follows,

\begin{equation}
\label{eq8} \left\{ \begin{array}{l}
 dT_{MC} = 2\left( {{\frac{{d\psi} }{{2\pi} }}} \right)^{2}{\frac{{(\Omega_{H} - \Omega _{D} )}}{{dZ_{H}} }},\\\\
 dP_{MC} = 2\left( {{\frac{{d\psi} }{{2\pi} }}} \right)^{2}{\frac{{\Omega_{D} \cdot (\Omega _{H} - \Omega _{D} )}}{{dZ_{H}} }}.
\end{array} \right.
\end{equation}

\noindent where $\Omega_{H}=a_{*}/(2\ r_{H})$ is the angular
velocity of the BH and
\begin{equation}
\label{eq9}
dZ_{H}=\left(\frac{R_{H}}{2\pi}\right)\frac{r_{H}^{2}+a^{2}\cos^{2}\theta}{(r_{H}^{2}+a^{2})\sin\theta}\left(\frac{d\theta}{dr}\right)dr
\end{equation}

\noindent In equations (8) and (9) $Z_{H}$ is the resistance of
the BH horizon with the surface resistivity given by $R_{H}=4\pi/c
\approx 377\ ohm$.

Incorporating the MC effects with the conservation of energy and
angular momentum, we have the following relations:

\begin{equation}
\label{eq10} \left\{ \begin{array}{l}
 {{\frac{{d}}{{dr}}}(\dot {M}_{D} L^{\dag}  - g_{MC} ) = 4\pi r(F_{total}
L^{\dag}  - H),{\rm} {\rm} {\rm} }\\\\
{{\frac{{d}}{{dr}}}(\dot
{M}_{D} E^{\dag}  - g_{MC} \cdot \Omega _{D} ) = 4\pi r(F_{total}
E^{\dag}-H\Omega _{D})}
\end{array} \right.
\end{equation}

\noindent where $H \equiv (1 / 4\pi r) \cdot dT_{MC} / dr$ is the
angular momentum flux due to the MC effects, and $g_{MC} $ is the
interior viscous torque of the disc in the MC process.

Thus we can derive the radiation flux and the total luminosity of
a thin disc by resolving equation (10).

\begin{equation}
\label{eq11}
\begin{array}{l}
F_{total} = {\frac{{1}}{{4\pi r}}}{\frac{{ - d\Omega _{D} /
dr}}{{(E^{\dag } - \Omega _{D} L^{\dag} )^{2}}}}{\int_{r_{in}}
^{r} {(E^{\dag}  - \Omega _{D} L^{\dag} )}}\times \\\\
\quad \quad  (\dot{M}_{D} \cdot dL^{\dag}  / dr + 4\pi rH)dr
\equiv F_{DA} + F_{MC}
\end{array}
\end{equation}

\begin{equation}
\label{eq12}
 g_{MC} = {\frac{{E^{\dag}  - \Omega _{D} L^{\dag
}}}{{ - d\Omega _{D} / dr}}}4\pi r \cdot F_{total}
\end{equation}

\begin{equation}
\label{eq13} {\cal L}_{total} = \dot {M}_{D} (1 - E_{in}^{\dag} )
+ 4\pi {\int_{r_{in}} ^{\infty} {H\Omega _{D}} } r \cdot dr
\end{equation}

As the magnetic field on the BH is supported by the surrounding
disc, there is some relation between $B_{H} $ and $\dot {M}_{D} $.
As a matter of fact the relation might be rather complicated, and
would be very different in different situations. One possibility
has been suggested by Moderski, Sikora \& Lasota (1997) and is
based upon the balance between the pressure of the magnetic field
on the horizon and the ram pressure of the innermost parts of an
accretion flow, i.e.,

\begin{equation}
\label{eq14} {{B_{H}^{2}} \mathord{\left/ {\vphantom {{B_{H}^{2}}
{\left( {8\pi}  \right)}}} \right. \kern-\nulldelimiterspace}
{\left( {8\pi}  \right)}} = P_{ram} \sim \rho c^{2}\sim {{\dot
{M}_{D}} \mathord{\left/ {\vphantom {{\dot {M}_{D}} {\left( {4\pi
r_{H}^{2}}  \right)}}} \right. \kern-\nulldelimiterspace} {\left(
{4\pi r_{H}^{2}} \right)}},
\end{equation}

Considering that equation (14) is not a certain relation between
$B_{H} $ and $\dot {M}_{D} $, we rewrite it as follows,

\begin{equation}
\label{eq15} B_{H}^{} \equiv \sqrt {2\kappa_{m}
\dot{M}_{D}/r_{H}^{2}} ,
\end{equation}

\noindent where $\kappa _{m} $ is a parameter indicating the
relative importance of the MC process with respect to the disc
accretion. The MC process dominates over disc accretion for
$\kappa _{m} \gg 1$, and it is dominated by the latter for $\kappa
_{m} \ll 1$. We emphasize that equation (15) is independent of
(14), one can always fix $\kappa _{m} $ with any given relation
between $B_{H} $ and $\dot {M}_{D} $.

As shown in Figure 4a, the total radiation flux from the disc with
the inner edge within $r_{ISCO} $ ($\lambda = 0.974$) can be
positive, though more energy and angular momentum are needed to
keep the Keplerian orbits of the accreting matter within $r_{ISCO}
$. The reason is that the energy and angular momentum are
transferred from the BH to the disc for $a_{ *}  > 0.3594$
(Blandford 1999), and the positive contribution due to the MC
effects exceeds the negative one due to the disc accretion. The BH
spin $a_{ *}  = 0.3594$ corresponds to $\Omega _{H} $ equal to
$\Omega _{D} $ at ISCO.

On the other hand, as shown in Figure 4b, the total radiation flux
from the inner disc could be negative for $a_{*}<0.3594$, giving
rise to an unphysical radiant region, although the inner edge is
located at ISCO with $\lambda = 1.000$. This result arises from
the magnetic transfer direction of the energy and angular momentum
from the inner disc to the BH for $a_{ *}  < 0.3594$.

From the above discussion we find that the MC process is a very
efficient mechanism in transferring energy and angular momentum
between the BH and its surrounding disc. The inner region of the
disc would be eventually disrupted for $a_{ *}  < 0.3594$, because
too much energy and angular momentum are transferred to the BH via
the MC process. Thus we infer that the inner edge will move
outward beyond ISCO. On the other hand, the inner edge can be
extended within ISCO for $a_{ *}  > 0.3594$, because the excess
energy and angular momentum can be transferred magnetically from
the rotating BH to the inner disc via the MC process.\\

From equation (12), we infer that $g_{MC} $ has the same sign as
the total radiation flux. Combining the MC effects with the fact
that an accretion disc cannot sustain a negative viscous torque,
we suggest that a reasonable inner edge should be constrained by a
positive radiation flux. Just like the case of SAD, one can always
find some reasonable radii for the inner edge of the disc, as long
as they are far away enough from the BH, to keep the radiation
flux positive throughout the whole disc. And what we have to do is
to find the smallest radius as the inner edge. Following Wang
(1995), we consider that the inner edge of the disc is the
position where the transportation of energy and angular momentum
on the disc just begins to exceed the adjustable range of the
interior viscous torque. However, this criterion can hardly be
formulated accurately. Alternatively, we suggest that the
criterion for the inner edge of a thin disc with a smoothly
distributed exterior torque can be written as

\begin{equation}
\label{eq16}
 \left( {{\frac{{dF_{total}} }{{dr}}}}
\right)_{r_{in}^{ +} }  = 0, \quad
\left({{\frac{{d^{2}F_{total}}}{{dr^{2}}}}} \right)_{r_{in}^{ +
}}> 0
\end{equation}

\begin{figure}
\vspace{0.5cm}
\begin{center}
\includegraphics[width=6cm]{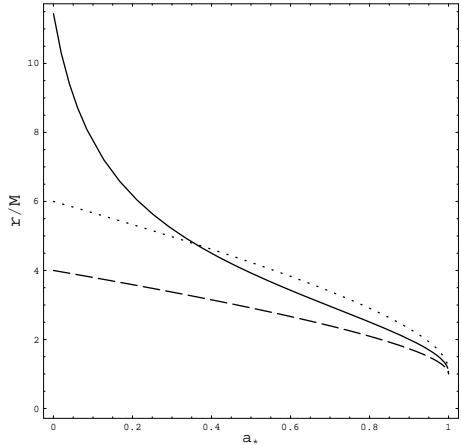}
\caption{Curves of the radius of the inner edge (solid line), the
radius of ISCO (dotted line) and the radius of the innermost bound
orbit (dashed line) versus the BH spin with $n = 6$ and $\kappa
_{m} = 1$.} \label{fig5}
\end{center}
\end{figure}

Incorporating equations (11) and (16), we have the radius of the
inner edge of the disc, ${{r_{in}}  \mathord{\left/ {\vphantom
{{r_{in} } {M}}} \right. \kern-\nulldelimiterspace} {M}}$, varying
with the BH spin $a_{ *}$ as shown in Figure 5. Inspecting Figure
5, we find that the deviation of $r_{in} $ from $r_{ISCO} $
depends on the BH spin as follows: (1) $r_{in} > r_{ISCO} $ for
$a_{\ast}  < 0.3594$, (2) $r_{in} < r_{ISCO} $ for $a_{\ast}
> 0.3594$ and (3) $r_{in} = r_{ISCO} $ at $a_{\ast}  = 0.3594$.

Based on the criterion (16), we can evaluate the influence of the
parameter $\kappa _{m} $ on the deviation of $r_{in} $ from
$r_{ISCO} $: (1) the turning point of the deviation $a_{\ast}  =
0.3594$ is independent of the value of $\kappa _{m} $; (2) the
greater $\kappa _{m} $ corresponds to the greater MC effects, and
thus to the greater deviation, and (3) we have $r_{in} = r_{ISCO}
$ for $\kappa _{m} = 0$, which corresponds to SAD without MC
effects according to equation (15).\\

Inspecting equations (11) and (12), we find that the MC torque
always vanishes at the inner edge, implying that the MC of the BH
with a thin disc does not disrupt the ¡°no torque boundary
condition¡±. However, there is an exception that the closed field
lines concentrate at the inner edge of the disc, where the MC
torque related to $g_{MC}$ at $r_{in}$ can be written in the form

\begin{equation}
\label{eq17} \left\{\begin{array}{l}
 dT_{MC} / dr \equiv 4\pi rH = \left( {g_{MC}}  \right)_{in} \cdot \delta
(r - r_{in}),\\
 \left( {g_{MC}}\right)_{in} = \tau _{0} \cdot
sign(\Omega _{H} - \Omega _{in} ).
\end{array} \right.
\end{equation}

\noindent This situation has been discussed in L04, where the MC
torque does not vanish at the inner edge of the disc. For
$a_{\ast} < 0.3594$, the criterion for the inner edge can be
equivalently written as

\begin{equation}
\label{eq18} \Omega _{H}(a_{*}) = \Omega _{D} (r_{in} ,a_{*}).
\end{equation}

Equation (18) implies that the inner edge is located at the radius
where the disc matter has the same angular velocity as the BH,
which is consistent with the conclusion in L04.

\begin{figure}
\vspace{0.5cm}
\begin{center}
\includegraphics[width=6cm]{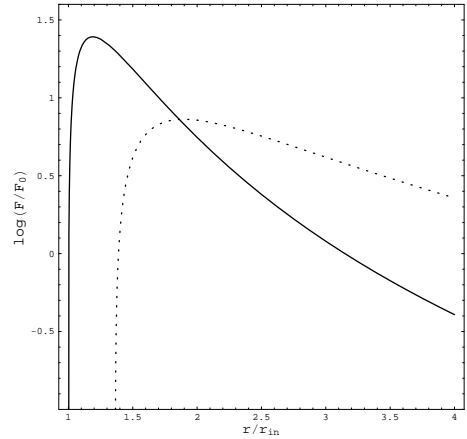}
\caption{ The curves of $F_{DA}/ F_{0}$(dotted line) and
$F_{MC}/F_{0}$ (solid line) versus $r/r_{in}$ for $a_{*}$= 0.75,
$n=6$ and $\kappa _{m}= 1.0$.} \label{fig6}
\end{center}
\end{figure}

\section{ MC EFFECTS ON DISC RADIATION}

The MC process has a significant effect on the disc radiation.
Compared with SAD, the radiation flux is much more concentrated at
the inner disc. By using equations (1) and (11) we have the curves
of $F_{DA}/F_{0}$ and $F_{MC}/F_{0}$ versus $r/r_{in}$ as shown in
Figure 6.

From equation (11) we obtain a very steep emissivity index up to
4.3$\sim $5.0 as shown in Figure 7, where the emissivity index is
defined as $\alpha \equiv - d\ln F_{total}/d\ln r$. This result is
consistent with the \textit{XMM-Newton} observation of the nearby
bright Seyfert 1 galaxy MCG-6-30-15 (Wilms J. et al. 2001;
Branduardi-Raymont et al. 2001).

\begin{figure}
\vspace{0.5cm}
\begin{center}
\includegraphics[width=6.0cm]{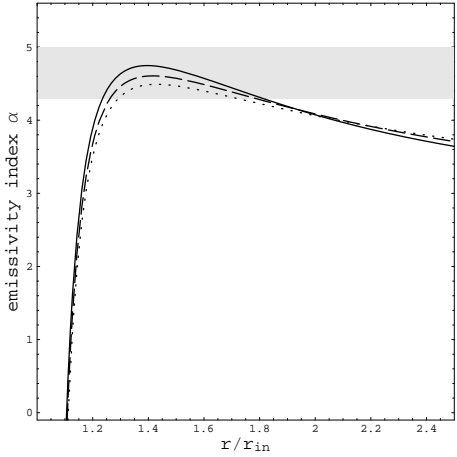}
\caption{ The curves of $\alpha \equiv - d\ln F_{total}/d\ln r$
versus $r/r_{in}$ for $a_{*} = 0.998$ and $\kappa _{m} = 10$. The
dotted, dashed and solid lines correspond to $n = $6, 7 and 8,
respectively. The emissivity index of the Seyfert 1 galaxy
MCG-6-30-15 inferred from the observation of XMM-Newton is shown
by the shaded region. } \label{fig7}
\end{center}
\end{figure}

Zhang et al. (1997, hereafter Z97) firstly invented an approach to
measure the BH spin by determining the radius of ISCO in fitting
the spectrum of the X-ray continuum from a thin relativistic disc.
Considering the deviation of the inner edge from ISCO due to the
MC effects, we modify equation (3) in Z97 as follows,

\begin{equation}
\label{eq19} r_{in} = \eta D{\left[ {{\frac{{F_{earth}} }{{2\sigma
g(\theta ,a_{\ast} ) \cdot Q(a_{\ast}  )}}}} \right]}^{1 /
2}{\left[ {{\frac{{f_{col} f_{GR} (\theta ,a_{\ast} )}}{{T_{col}}
}}} \right]}^{2}
\end{equation}

\noindent where $g(\theta ,a_{\ast}  )$ and $f_{GR} (\theta
,a_{\ast}  )$ have the same meaning given in Z97. We ignore the
effect of magnetic field on the geodesic of photons, so they are
only determined by the metric around the BH. In this paper, we use
the values of $g(\theta ,a_{\ast}  )$ and $f_{GR} (\theta
,a_{\ast} )$ from Table 1 in Z97. $\eta \equiv r_{in} / r_{peak}$.
And the other quantities are defined as follows:

\begin{equation}
\label{eq20}
\begin{array}{l}
F_{earth} = g(\theta ,a_{\ast} ){\cal L}_{total} / 2\pi D^{2},\\
T_{col} = f_{GR} (\theta ,a_{\ast} )f_{col} \cdot T_{peak}
\end{array}
\end{equation}

Compared with Z97, the MC effects are taken into account in our
model. In addition, we introduce a factor $Q(a_{\ast}  )$ to
modify the approximate relation between the bolometric luminosity
of the disc and the peak emission region (Makishima et al. 1986)
as follows,

\begin{equation}
\label{eq21} {\cal L}_{total} \equiv 4\pi \sigma r_{peak}^{2}
T_{peak}^{4} \cdot Q(a_{\ast}  )
\end{equation}

The MC effects on the data fitting behave at least two aspects:
(1) the inner edge deviates from ISCO, and (2) the radius
$r_{peak} $ of the peak value of the radiation flux is much closer
to the inner edge. Incorporating equations (19)---(21) with the
criterion (16), we have the BH spins of GRO J1655-40 and GRS
1915+105 as shown in Table 1.

\begin{table*}
\caption{Measuring BH spins of GRO J1655-40 and GRS 1915+105 based
on X-ray continuum with the MC effects ($n$=6.0 and
$\kappa_{m}$=1.0).}
\begin{center}
\begin{tabular}
{|c|c|c|c|c|c|} \hline
\raisebox{-1.5ex}[0pt]{Source}&\multicolumn{4}{|c|}{Observed
parameters}&\raisebox{-1.5ex}[0pt]{BH spin}\\
\cline{2-5} &$F_{earth}$($erg\cdot cm^{-2}\cdot s^{-1}$)&
$T_{col}$($k/Kev$)& $\theta$(deg)& M($M_{\odot}$)\\
\hline GRO J1655-40&$3.3\times10^{-8}$&1.36& 70& $6.0\sim 6.6$& $0.932\sim 0.960$\\
\hline GRS 1915+105&$4.4\times10^{-8}$&2.27& 70& $10\sim 18$&$0.927\sim 0.996$ \\
\hline
\end{tabular}
\end{center}
\end{table*}

In Z97, the factor $Q(a_{*})$ is set to unity, and it is argued
that the approximation is accurate to within 10\% for a wide range
of the model parameter space. In this paper, we check the relation
in equation (21) carefully and find that the approximation made in
Z97 works well only for the non-relativistic discs. Once the
relativistic effects are taken into account, the correction from
$Q(a_{\ast}  )$ would deviate unity significantly by a factor of
several orders. On the other hand, the correction above in our
paper is also partially from the MC effects. We emphasize that the
results in our paper contains fully relativistic effects, which is
different from the ones in Z97. And one can also find that the
derived spins of the black holes in Z97 would be much smaller if
the correction $Q(a_{*})$ is considered seriously.

According to the observation, jets appear both in GRO J1655-40 and
GRS 1915+105. As shown in Figure 3, two kinds of magnetic field
configurations are contained in our model. The closed field lines
connecting the BH and the disc correspond to the MC process. The
open field lines connecting the BH to the remote astrophysical
load correspond to BZ process, which is widely used to interpret
the jet-production in X-ray binaries, AGNs and GRBs (Blandford \&
Znajek 1977, Blandford 1999). Based on the argument given in W03
we find that the BZ process can coexist with the MC process for
the parameters in Table 1. Thus the jet-production in GRO J1655-40
and GRS 1915+105 can be interpreted naturally based on this model.

\section{CONCLUSION AND DISCUSSION}

In this paper the MC effects on the radiation from a relativistic
thin disc are discussed, and some issues related to this toy model
are addressed as follows.

(1) As argued by Uzdensky (2005), the MC of a rotating BH with its
surrounding disc can be regarded as a stable magnetic connection,
which is dramatically different from the magnetic connection of a
rotating neutron star with a disc. Since a BH does not have a
conducting surface, the magnetic field lines frozen into a
rotating conducting disk can slip on the horizon. Compared with
the BH, a neutron stars is a highly-conducting star. Therefore,
each field line connecting the star to the disk is subject to a
continuous twisting, and no steady magnetic connection exists in
the case of a rotating neutron star with a disc (Ghosh, Lamb \&
Pethick 1977, Ghosh \& Lamb 1979a, b; Eksi, Herquist \& Narayan
2005).

(2) The MC discussed in this paper involves a large-scale magnetic
field connecting a rotation BH with a thin disc. As argued in L02,
this type of MC is very different from those involving a
small-scale magnetic field connecting the matter inside ISCO with
the matter outside ISCO (Krolik 1999; Gammie 1999; Agol \& Krolik
2000), and the ``no torque boundary condition'' is assumed based
on the consideration of this MC with a thin disc.\\

The contribution of this paper lies in the MC effects on disc
radiation based on the deviation of the inner edge from ISCO, and
the main results are summarized as follows.

It is shown that the inner edge of SAD should be located at or
outside ISCO, which is regarded as the inner boundary of SAD. ISCO
is the smallest one among all the possible stable circular orbits
in SAD, although it is not compelling to taken ISCO as the inner
edge of the disc.

Considering the magnetic transfer of energy and angular momentum
between a rotating BH and its surrounding accretion disc, we
propose a criterion for the position of the inner edge based on a
reasonable constraint to the radiation flux from the disc. It
turns out that the deviation of the inner edge from ISCO depends
on the direction of the magnetic transfer, which is eventually
determined by the BH spin: the radius $r_{in} $ could be greater
and less than $r_{ISCO} $ for $a_{ *}  $ being less and greater
than 0.3594, respectively.

In addition, the MC effects on the boundary condition of the disc
are discussed. It is argued that the ``no torque boundary
condition'' is not affected at the presence of the MC process
except that the magnetic field is concentrated completely at the
inner edge. Only in this extreme case the MC effects can be
treated effectively as a ``non-zero boundary torque''.

It has been argued that no stable geodesic circular orbits exist
within ISCO (B72). The conclusion is derived based on geodesic of
a test particle in the Kerr metric. It is worth noting that the
orbits of the accreting matter cannot be regarded as geodesics
because of the interior viscous torque and the presence of a
strong magnetic field related to the MC process, and the result
that $r_{in} $ is less than $r_{ISCO} $ for a disc around a
fast-rotating BH seems not in
conflict with the conclusion given in B72.\\

\noindent \textbf{Acknowledgments.} This work is supported by the
National Natural Science Foundation of China under Grant Numbers
10573006 and 10121503. The anonymous referee is thanked for his
(her) helpful comments and suggestions.

\end{document}